# Vapor Drainage in the Protolunar Disk as the Cause for the Depletion in Volatile Elements of the Moon

Nicole X. Nie[1*], Nicolas Dauphas[1]

[1]Origins Laboratory, Department of the Geophysical Sciences and Enrico Fermi Institute, The University of Chicago, 5734 South Ellis Avenue, Chicago IL 60637, USA

[*]To whom correspondence should be addressed: xike@uchicago.edu

**Abstract**

Lunar rocks are severely depleted in moderately volatile elements such as Rb, K, and Zn relative to Earth. Identifying the cause of this depletion is important for understanding how the Earth-Moon system evolved in the aftermath of the Moon-forming giant impact. We measured the Rb isotopic compositions of lunar and terrestrial rocks to understand why moderately volatile elements are depleted in the Moon. Combining our new measurements with previous data reveals that the Moon has an $^{87}$Rb/$^{85}$Rb ratio higher than Earth by +0.16±0.04 ‰. This isotopic composition is consistent with evaporation of Rb into a vapor medium that was ~99% saturated. Evaporation under this saturation can also explain the previously documented isotopic fractionations of K, Ga, Cu and Zn of lunar rocks relative to Earth. We show that a possible setting for achieving the same saturation upon evaporation of elements with such diverse volatilities is through viscous drainage of a partially vaporized protolunar disk onto Earth. In the framework of an α-disk model, the α-viscosity needed to explain the ~99% saturation calculated here is $10^{-3}$ to $10^{-2}$, which is consistent with a vapor disk where viscosity is controlled by magnetorotational instability.

1. Introduction

The prevailing view of the origin of the Moon is that it formed by a collision between the protoEarth and a large impactor that has been named Theia, which produced a disk of liquid and vapor from which the Moon grew (Hartmann & Davis 1975; Cameron & Ward 1976). Simulations of this giant impact can explain the angular momentum of the Earth-Moon system and the small mass of the lunar core but other chemical and isotopic constraints are more challenging to reproduce (Canup & Asphaug 2001; Pahlevan & Stevenson 2007; Canup 2012; Cuk & Stewart 2012; Reufer et al. 2012; Nakajima & Stevenson 2015; Lock et al. 2018). Among these is the lunar depletion in moderately volatile elements (MVEs) such as Rb, K, and Zn (Ringwood & Kesson 1977; O'Neill 1991; Albarède et al. 2015; Righter 2019). The settings considered to explain these depletions are (*i*) incomplete condensation of a vapor cloud surrounding the Earth (synestia) in the aftermath of a very energetic impact event (Lock et al. 2018), (*ii*) accretion of the Moon from a partially condensed, volatile-depleted silicate melt carried across the Roche limit by viscous spreading (Canup et al. 2015), (*iii*) removal of volatile elements from a stratified protolunar disk by accretion of the vapor layer onto Earth (Charnoz & Michaut 2015; Wang et al. 2019), and (*iv*) loss to space from the lunar magma ocean (Day & Moynier 2014; Saxena et al. 2017). One difficulty in assessing these scenarios is the lack of thermodynamic data constraining the condensation/evaporation behaviors of MVEs at the very high temperatures (presumably >3500 K; Canup et al. 2015; Lock et al. 2018) involved in the formation of the Moon. Isotopic analyses of several MVEs have revealed heavy isotope enrichments of lunar rocks compared to Earth (Herzog et al. 2009; Paniello et al. 2012; Kato et al. 2015; Wang & Jacobsen 2016; Kato & Moynier 2017; Pringle & Moynier 2017; Wimpenny et al. 2019) but the extents of those fractionations are uncertain, and a quantitative understanding of how they relate to the conditions of the Moon formation is still missing.

We report here new and high-precision Rb isotopic analyses of lunar and terrestrial samples that we use to understand why the Moon is depleted in MVEs relative to Earth. In Sect. 2, we report the results of Rb isotopic analyses of Apollo lunar samples (primarily basalts) indicating that the Moon is enriched in the heavy isotopes of Rb relative to Earth. In Sect. 3, we calculate the saturation of Rb (~99 %) in the vapor medium needed to explain the heavy Rb isotope enrichment of lunar rocks by evaporation, and show that the same degree of saturation can account for previously documented heavy isotope enrichments in other MVEs K, Ga, Cu, and Zn. In Sect. 4, we show that this saturation can be explained in the context of viscous drainage of vapor in the protolunar disk onto Earth if the $\alpha$-viscosity of the vapor layer was $10^{-3}$ to $10^{-2}$, which is consistent with viscosities predicted if the magnetorotational instability (MRI) was active.

2. Samples and Results

We have measured the isotopic compositions of Rb in six Apollo lunar samples comprising five mare basalts and one norite (Table 1) to better constrain the bulk Rb isotopic composition of the Moon, which is expressed in $\delta^{87}$Rb notation (the permil departure of the $^{87}$Rb/$^{85}$Rb ratio relative to reference material NIST SRM984). To compare to the Earth, we have also measured the Rb isotopic compositions of terrestrial basalts and granites (Table 1). Pringle & Moynier (2017) previously measured the Rb isotopic compositions of lunar rocks and found a barely resolvable difference in $\delta^{87}$Rb of +0.17±0.13 ‰ between the Moon and Earth using a protocol that presumably did not fully separate Rb from K (see the Appendix). The motivations for our new Rb isotopic analyses were therefore to improve on the precision of these analyses and test their accuracy using a protocol that fully separated K from Rb (see the Appendix for details). Basalt samples have been prioritized because the chemical and isotopic compositions of other lithologies could have been more severely affected by lunar magma ocean differentiation. Indeed, Rb is incompatible during magmatic processes but it can partition in feldspar. It is thus conceivable that it would have been fractionated isotopically when an anorthositic flotation crust formed from the lunar magma ocean. For example, the cataclastic norite sample 77215 from Apollo 17 measured in this study, which is a white-colored fragment containing abundant plagioclase, shows a fractionated Rb isotopic composition (-0.15 ± 0.03 ‰) compared to mare basalts (average +0.03 ± 0.03 ‰). All basalts have similar $\delta^{87}$Rb and $\delta^{41}$K values, so it is unlikely that their isotopic compositions were affected by Rb and K volatilization during eruption at the lunar surface.

In the following, we compare Rb with K as these are two alkali elements with similar geochemical behaviors (O'Neill 1991). To mitigate the effect of magmatic fractionation, we plot the $\delta^{87}$Rb (this study and Pringle & Moynier 2017) and $\delta^{41}$K (Wang & Jacobsen 2016) values of lunar and terrestrial rocks against the La/U ratios. The motivation for doing so is that La and U have a similar geochemical behavior to Rb and K during magmatic processes (O'Neill 1991) but unlike the latter, La and U are two refractory elements that were immune to volatile-loss processes and should therefore be in chondritic proportion in the bulk Moon and Earth. Samples with near-chondritic La/U ratios (~27.6) are therefore less likely than others to have had their Rb and K isotopic compositions fractionated by magmatic differentiation. Focusing on samples with La/U ratios within 0.5× to 1.5× the chondritic value, we estimate that the Moon and Earth differ in their $\delta^{87}$Rb and $\delta^{41}$K values by +0.16 ± 0.06 and +0.40 ± 0.05 ‰, respectively.

An alternative approach for estimating the Rb and K isotopic compositions of the Earth and Moon is to interpolate the trends between $\delta^{87}$Rb, $\delta^{41}$K, and La/U ratios to the chondritic La/U ratio (Fig. 1). Variations in the La/U ratio most likely reflect magmatic differentiation processes (including mixing with a KREEP component characterized by (La/U)/(La/U)$_{CI}$ ≃ 0.65; Warren 1989), and we take the $\delta^{87}$Rb and $\delta^{41}$K values at (La/U)/(La/U)$_{CI}$ = 1 (unfractionated La/U ratio) as representative of the bulk Moon. Accordingly, we estimate that the $\delta^{87}$Rb and $\delta^{41}$K values of the Moon are heavier than those of the Earth by +0.16 ± 0.04 ‰ and +0.41 ± 0.07 ‰, respectively, which agrees with the values given above by taking the average of all samples with near-chondritic La/U ratios. The slightly positive slope of lunar $\delta^{87}$Rb values against La/U ratios is controlled by the two heaviest basalts but excluding them from the regression does not change the estimated bulk lunar $\delta^{87}$Rb value.

3. Vapor Saturation Conditions for the Lunar Depletion of MVEs

In some models of lunar formation (scenarios (*i*) and (*ii*) above in Section 1), the depletions in Rb and K are interpreted to reflect incomplete condensation (Canup et al. 2015; Lock et al. 2018). However, this process cannot explain the heavy isotopic compositions of Rb and K in the Moon because at the temperature of ~3500 K considered in these models, equilibrium isotopic fractionation between vapor (as Rb and K atoms) and condensate is too small to account for the observed values (Dauphas et al. 2018a), and kinetic isotopic fractionation would enrich the partial condensate (the Moon) in the light isotopes (Richter 2004), which is opposite to what is observed. Most likely, the elevated $\delta^{87}$Rb and $\delta^{41}$K values of the Moon relative to Earth reflect partial evaporation. A model of evaporation from the lunar magma ocean (scenario *iv* above) indicates that for Na, whose volatility is similar to K, the cumulative loss over the lifetime of the magma ocean is small (5-20%; Saxena et al. 2017), which is insufficient to account for the six-fold depletion of K and Rb in the Moon relative to Earth. More work, however, is needed to fully assess this scenario, especially for the more volatile element Zn (Day & Moynier 2014; Charnoz et al. 2019; Young & Tang 2019).

Loss of MVEs could have happened through drainage and loss of vapor from the protolunar disk onto the protoEarth (scenario (*iii*) above; Charnoz & Michaut, 2015). In this model, it is assumed that the protolunar disk is stratified, comprising a liquid layer at the midplane surrounded above and below by a vapor layer. MVEs are evaporated from the liquid layer into the vapor layer. The fate of MVEs and their depletions in the Moon depend sensitively on the viscosity of the vapor layer. The reason is that viscosity mediates transport of mass and angular momentum across the disk. Little loss of liquid from the disk to the Earth could have taken place over much of the cooling history of the disk, because a gravitationally stable low-viscosity region would be present within ~1.7 $R_\oplus$ that would have acted as a barrier preventing liquid from being accreted by the Earth. If the viscosity of the vapor was low, it would have remained stagnant and no volatile loss would have occurred. If the viscosity of the vapor was high due to MRI (Carballido et al. 2016; Gammie et al. 2016), the vapor layer carrying MVEs would have been efficiently accreted by the Earth, leaving a disk depleted in volatile elements.

Drainage of the vapor layer onto Earth is a complex process that would have involved some evaporation from the liquid to the vapor layer (Charnoz & Michaut 2015). The kinetics and isotopic fractionation associated with such evaporation depend on the mode of transport inside (advective *vs*. diffusive), and across (equilibrium *vs*. kinetic) the liquid and vapor layers (Craig & Gordon 1965; Richter et al. 2002; 2007; Richter 2004; Dauphas et al. 2015). Given the rate of heat loss experienced by the disk through radiation to space, much of the vapor layer would have convected vigorously (Thompson & Stevenson 1988; Ward 2011). Furthermore, given the difference in orbital velocities between the vapor and liquid layers, it is likely that Kelvin–Helmholtz instabilities developed (Thompson & Stevenson 1988; Charnoz & Michaut 2015) that prevented the development of a viscous layer at the vapor–liquid interface. In this context, where transport within the liquid and vapor layers is entirely advective, the extent of isotopic fractionation of element $i$ during evaporation depends (through the Hertz–Knudsen equation) on its vapor saturation (Richter et al. 2002; 2007; Dauphas et al. 2015), expressed as $S_i = P_i/P_{i,\text{eq}}$ where $P_i$ is the partial vapor pressure of element $i$ and $P_{i,\text{eq}}$ is its equilibrium vapor pressure at the relevant temperature.

If $S_i = 1$, there is no net evaporative flux, the vapor is in thermodynamic equilibrium with the liquid, and the isotopic fractionation between vapor and liquid is entirely equilibrium in nature (noted $\Delta_{eq,i}^{v-l}$ hereafter). *Ab initio* calculations give the equilibrium fractionation between monoatomic K vapor (the dominant gas species in conditions relevant to lunar formation) and K-feldspar (taken as a proxy for K in silicate melt) to be $\Delta_{eq,K}^{v-l} \simeq -0.012$ to $-0.024$‰ at 3500 K and –0.055 to –0.107‰ at 1650 K (Dauphas et al. 2018a; Zeng et al. 2018; Li et al. 2019; Zeng et al. 2019). Note that in silicate melts with (K, Na)/Al ratios of <1 relevant to magmas of bulk silicate Earth or Moon compositions, K and Rb charge balance $Al^{3+}$ in tetrahedral coordination, so to first order feldspar is a reasonable model structure for those melts (Seifert et al. 1982; Toplis et al. 1997). Rubidium forms chemical bonds of similar strengths to K (Zeng et al. 2019), and we estimate that the vapor–liquid equilibrium fractionation for Rb should be $\Delta_{eq,Rb}^{v-l} \simeq -0.003$ to $\simeq -0.007$‰ at 3500 K and –0.016 to –0.031‰ at 1650 K ($\Delta_{eq,Rb}^{v-l}/\Delta_{eq,K}^{v-l} \simeq 0.3$).

If $S_i = 0$, the flux is unidirectional (from the liquid to the vapor) and the isotopic fractionation is mostly kinetic in nature (noted as $\Delta_{kin,i}^{v-l}$ hereafter). Experiments of K evaporation in vacuum give a kinetic isotopic fractionation factor $\Delta_{kin,K}^{v-l} = [(39/41)^{0.43} - 1] \times 1000 = -22$ ‰ (Richter et al. 2011). No data are available for Rb evaporation, but we can reasonably assume that it follows the same mass dependence, which gives $\Delta_{kin,Rb}^{v-l} = [(85/87)^{0.43} - 1] \times 1000 = -10$ ‰.

If evaporation took place in a medium that was partially saturated ($0 < S_i < 1$), the instantaneous isotopic fractionation factor between vapor and liquid ($\Delta_i^{v-l}$) would be intermediate between the equilibrium and kinetic values given above (Dauphas et al. 2015):

$$\Delta_i^{v-l} = \Delta_{eq,i}^{v-l} + (1 - S_i)\Delta_{kin,i}^{v-l}. \qquad (1)$$

This expression gives the instantaneous isotopic fractionation between vapor and liquid. In the model of Charnoz & Michaut (2015) (scenario *(iii)* above), the vapor is continuously removed by accretion onto Earth through viscous spreading, leaving behind an MVE-depleted liquid layer. This is a complex process that to first order can be modeled using a Rayleigh distillation:

$$\delta_{i,l} - \delta_{i,0} \simeq \left[\Delta_{eq,i}^{v-l} + (1 - S_i)\Delta_{kin,i}^{v-l}\right] \ln f_i, \qquad (2)$$

where $\delta_{i,l}$ is the isotopic composition (in ‰) of the residual liquid after evaporation, $\delta_{i,0}$ is the initial isotopic composition, and $f_i$ is the fraction of the element remaining after vapor loss. In the context of lunar formation, the present isotopic composition of the Earth would be the starting composition (because in the Earth–Moon system, most of Rb and K resides in Earth), the lunar composition would be the residual liquid, and the fraction of either Rb or K remaining would be ~0.17 (corresponding to a six-fold depletion; Ringwood & Kesson 1977; O'Neill 1991; Albarède et al. 2015). Wang et al. (2019) argued that the heavy K isotopic composition of the Moon relative to Earth could be explained through evaporative loss where each step in the distillation involves vapor–liquid equilibrium. Using *ab initio* data (Dauphas et al. 2018b; Zeng et al. 2018; Li et al. 2019; Zeng et al. 2019), we calculate that the $\delta^{41}K$ value of the residue remaining after 83% loss of K by distillation under equilibrium conditions ($S_i = 1$ in Eq. 2) should be +0.021 to +0.042 ‰ at 3500 K (Canup et al. 2015; Lock et al. 2018) and +0.097 to

+0.190‰ at 1650 K (Wang et al. 2019). These values are much smaller than the observed Moon–Earth difference of +0.41 ± 0.07 ‰. We therefore disagree with the assessment of Wang et al. (2019) that the heavy K isotopic composition of the Moon (and by extension Rb and possibly Zn) can be explained by evaporative loss under equilibrium conditions. The data instead call for loss in an undersaturated medium involving kinetic isotopic fractionation.

Using Eq. 2, we calculate the vapor saturation needed to explain the measured isotopic fractionations between lunar and terrestrial rocks for Rb and K. We find values of $S_{\rm Rb} = 0.991$ and $S_{\rm K} = 0.990$. If evaporation had occurred in a more undersaturated medium, kinetic isotope effects would have left the Moon more isotopically fractionated in Rb and K than measured. Conversely, if the evaporation had occurred in a medium closer to saturation, the instantaneous fractionation would approach the equilibrium value, leaving the Moon less isotopically fractionated than it is.

In Fig. 2, we test if this calculated level of saturation could also explain the isotopic fractionations documented for other MVEs. Given the very high temperature involved, we can neglect the equilibrium term in Eq. 2 and to a good approximation, we have

$$\delta_{i,l} - \delta_{i,0} \simeq (1 - S_i)\Delta^{v-l}_{kin,i} \ln f_i. \qquad (3)$$

If all MVEs experienced evaporation under the same saturation, we would expect to find a linear relationship between $\delta_{\rm Moon} - \delta_{\rm Earth}$ and $\Delta^{v-l}_{kin,i} \ln f_i$ whose slope is $1 - S$. The MVEs Rb, Ga, Cu, K, and Zn indeed define a straight line corresponding to $S = 0.989 \pm 0.002$ (Fig. 2; weighted linear regression). Refractory elements such as Ca (Simon & DePaolo 2010) and Ti (Millet et al. 2016) are not shown in Fig. 2 but they would plot at the origin (like Li, Mg, and Si) and would therefore follow the trend defined by MVEs. The only exception is Sn, which seems to be enriched in the lighter isotopes in lunar basalts compared to terrestrial rocks (Wang et al. 2019). This could be due to a large vapor–liquid equilibrium isotopic fractionation and a large evaporation coefficient for Sn (Sect. 4), such that the equilibrium term would dominate over the kinetic term in Eq. 2 $\left(\left|\Delta^{v-l}_{eq,i}\right| > \left|(1 - S_i)\Delta^{v-l}_{kin,i}\right|\right)$. It could also reflect the fact that the lunar and terrestrial Sn isotopic compositions are not well known, as this element can exist in several oxidation states in planets ($Sn^0$, $Sn^{2+}$, and $Sn^{4+}$), which can drive large isotopic fractionation (Badullovich et al. 2017; Dauphas et al. 2018a; Wang et al. 2018; Roskosz et al. 2019).

4. Viscous Drainage of Vapor in an MRI-active Protolunar Disk

The fact that elements showing great diversity in their volatilities (*e.g.*, Zn is ~200-fold depleted while K and Rb are ~6-fold depleted; Ringwood & Kesson 1977; O'Neill 1991; Albarède et al. 2015) experienced evaporation under similar saturation $S \simeq 0.99$ is a fundamental observation and critical test for scenarios of lunar volatile depletion. As demonstrated below, this is a natural outcome of the model of volatile loss by viscous–drainage of the vapor layer of the protolunar disk onto Earth. Still, further work is needed to evaluate whether other scenarios of Moon formation (Pahlevan & Stevenson 2007; Canup et al. 2015; Nakajima & Stevenson 2015; Saxena et al. 2017; Lock et al. 2018; Charnoz et al. 2019; Righter 2019; Young & Tang 2019) can similarly account for MVE loss by evaporation under a saturation of 99%.

The fact that the vapor was undersaturated means that there was a net evaporative flux from the liquid to the vapor layer, which must have been balanced by vapor removal. We consider a scenario (Fig. 3) where the disk is composed of a liquid layer at the midplane of the protolunar disk, overlain by a vapor layer that is drained efficiently to the Earth due to its high viscosity, presumably powered by MRI (Charnoz & Michaut 2015; Carballido et al. 2016; Gammie et al. 2016). The timescale for producing a vapor layer with a hydrostatic profile for element $i$ at saturation $S_i$ is obtained by dividing the total mass of an element in the vapor layer by the flux across the liquid/vapor interface given by the Hertz–Knudsen equation (see the Appendix),

$$t_{S,i} = \frac{\pi S_i}{\gamma_i(1-S_i)} \sqrt{\frac{m_i R^3}{\bar{m} G M_\oplus}}, \qquad (4)$$

where $\gamma_i$ is a dimensionless evaporation coefficient (available experimental results give values between 0.017 and 0.13 for evaporation of K from silicate melt; Fedkin et al. 2006; Richter et al. 2011); $m_i$ and $\bar{m}$ are the mass of the vapor species for element $i$ (e.g., ~39 g mol$^{-1}$ for K because monoatomic K dominates the vapor) and the mean molecular mass of the bulk vapor (~30 g mol$^{-1}$; Ward 2011; Charnoz & Michaut 2015), respectively; $R$ is the distance to Earth's center (~1.7 $R_\oplus$=11×10$^6$ m if we take the outer edge of the gravitationally stable liquid disk; Charnoz & Michaut 2015), and $M_\oplus$ is Earth's mass. Importantly, this timescale is independent of the volatility ($P_{i,\text{eq}}$) of the element considered (and the uncertainties attached to it at the elevated temperature considered) because $t_{S,i}$ is the ratio of the mass of an element in the vapor layer divided by the evaporation flux across the liquid/vapor boundary, which both scale linearly with $P_{i,\text{eq}}$. In our preferred scenario (Fig. 3), vapor removal would occur by accretion onto the Earth through viscous spreading, which takes place on a timescale

$$t_a = \frac{2R^2}{3\nu}, \qquad (5)$$

where $\nu$ is the kinematic viscosity (in m$^2$ s$^{-1}$).

Both timescales ($t_{S,i}$ and $t_a$) are independent of volatility ($P_{i,\text{eq}}$), meaning that to first order, the level of saturation in the vapor layer should be similar for all elements. More volatile elements would still be more efficiently partitioned into the vapor layer and more efficiently lost than less volatile elements, explaining the different extents of depletion of MVEs measured in lunar rocks. Some variations may still exist in the saturation level due to differences in atomic/molecular mass ($m_i$) and more importantly evaporation coefficients ($\gamma_i$), so further experimental work is needed to better characterize the latter.

If $t_a \ll t_{S,i}$, vapor is removed more rapidly than it can be replenished by vaporization from the liquid such that the saturation would be lower than $S \simeq 0.99$, and the isotopic fractionation would be larger than that measured in lunar rocks. Conversely, if $t_a \gg t_{S,i}$, the saturation would be higher than calculated, and the isotopic fractionation would be smaller than measured in lunar rocks. Equating these two timescales, we can calculate the viscosity needed to maintain the vapor at the saturation level dictated by isotopes

$$\nu = \frac{2\gamma_i(1-S_i)}{3\pi S_i} \sqrt{\frac{\bar{m} R G M_\oplus}{m_i}}, \qquad (6)$$

or alternatively the saturation set for each element by a given viscosity (we use the approximation $S_i \sim 1$)

$$S_i \simeq 1 - \frac{3\pi\nu}{2\gamma_i}\sqrt{\frac{m_i/\overline{m}}{RGM_\oplus}}. \qquad (7)$$

As shown in Fig. 2, most MVEs record the same saturation level $S \simeq 0.99$. In the context of our model, the only element-specific controls on the saturation level are the mass of the vapor species ($\sqrt{m_i/\overline{m}}$) and the evaporation coefficient ($\gamma_i$). The $\sqrt{m_i/\overline{m}}$ term varies little from one element to another (1.14 for K, 1.46 for Cu, 1.48 for Zn, 1.52 for Ga, 1.69 for Rb) but the similarity in saturation also requires that the evaporation coefficients $\gamma_i$ be similar (Eq. 7). As discussed in Sect. 3, the isotopic composition of Sn in lunar rocks seems to support evaporation under equilibrium conditions ($S = 1$; Wang et al. 2019). A possible explanation for this peculiar behavior is that the evaporation coefficient for Sn is much higher than for other MVEs (Eq. 7), which can be tested in the future by performing evaporation experiments.

To assess whether the calculated kinematic viscosity of the vapor layer (Eq. 6) is realistic or not, we put it in the context of an $\alpha$-disk, where the viscosity is expressed as $\nu = \alpha c^2/\Omega$, with $\alpha$ a dimensionless number, $c = \sqrt{kT/\overline{m}}$ the sound speed ($k$ is the Boltzmann constant), and $\Omega = \sqrt{GM_\oplus/R^3}$ the Keplerian angular velocity. We calculate that $\alpha$ must have been on the order of $10^{-3}$ to $10^{-2}$ in order to explain the inferred saturation of ~0.99. Carballido et al. (2016) and Gammie et al. (2016) estimated that in the protolunar disk, where a fraction of alkalis could have been ionized, MRI could have sustained an $\alpha$ value on the order of $10^{-2}$. Our estimate of $\alpha$ based on the isotopic compositions of Rb, K, and other volatile elements is thus entirely consistent with theoretical expectations, suggesting that MRI-powered viscous drainage is a viable mechanism for explaining the depletion of MVEs in the Moon. At the present time, we cannot tell whether such viscous drainage of vapor would have been a steady process occurring throughout the lifetime of the protolunar disk, or whether it was episodic.

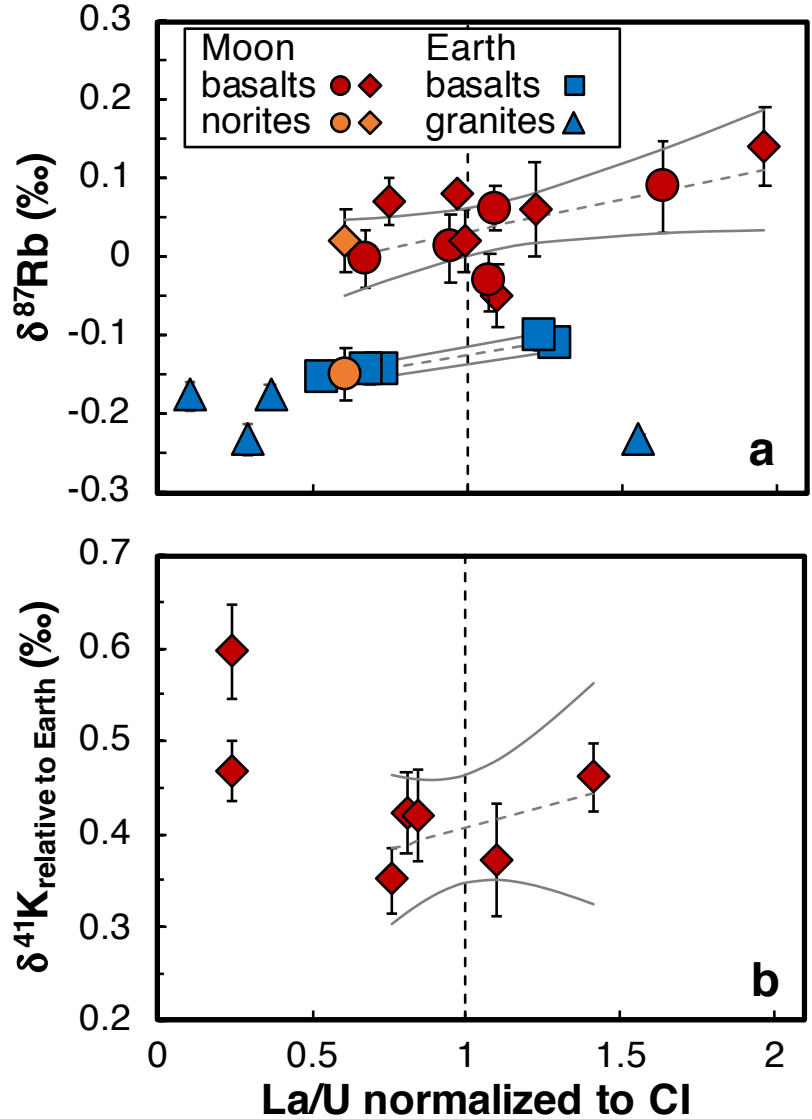

**Figure 1.** Rb and K isotopic compositions of lunar (red and orange circles and diamonds) and terrestrial (blue squares and triangles) samples, plotted against La/U ratios (taken here as an indicator of magmatic differentiation) normalized to the CI chondritic ratio (27.6). The isotopic compositions of the bulk Moon and bulk Earth were estimated using unweighted linear regressions interpolated to a CI chondrite-normalized La/U ratio of 1. (a) The $\delta^{87}Rb$ values of the bulk Moon and the bulk Earth are estimated to be +0.03 ± 0.03 and −0.13 ± 0.01 ‰ respectively, which corresponds to a Moon–Earth difference of +0.16 ± 0.04 ‰. The lunar $\delta^{87}Rb$ data are from this study (red circles=lunar basalts, orange circle=norite; Table 1) and from Pringle & Moynier (2017) (red diamonds=basalts, orange diamond=norite; Table 2), and the La and U concentrations are from the Lunar Sample Compendium (https://curator.jsc.nasa.gov/lunar/lsc/). The norite sample measured in this study (orange circle) was excluded from the regression due to its very light Rb isotopic composition (this sample is heterogeneous and the fragment measured was a white-colored piece containing abundant plagioclase, which could have fractionated Rb isotopic composition). The terrestrial $\delta^{87}Rb$ values and the La/U ratios can be found in Table 1. (b) The bulk Moon $\delta^{41}K$ value is estimated to be +0.41±0.07 ‰ relative to the Earth. The two fractionated K isotope data points are breccias with granitic components. The $\delta^{41}K$ values of lunar samples relative to the Earth are from Wang & Jacobsen (2016) (Table 2).

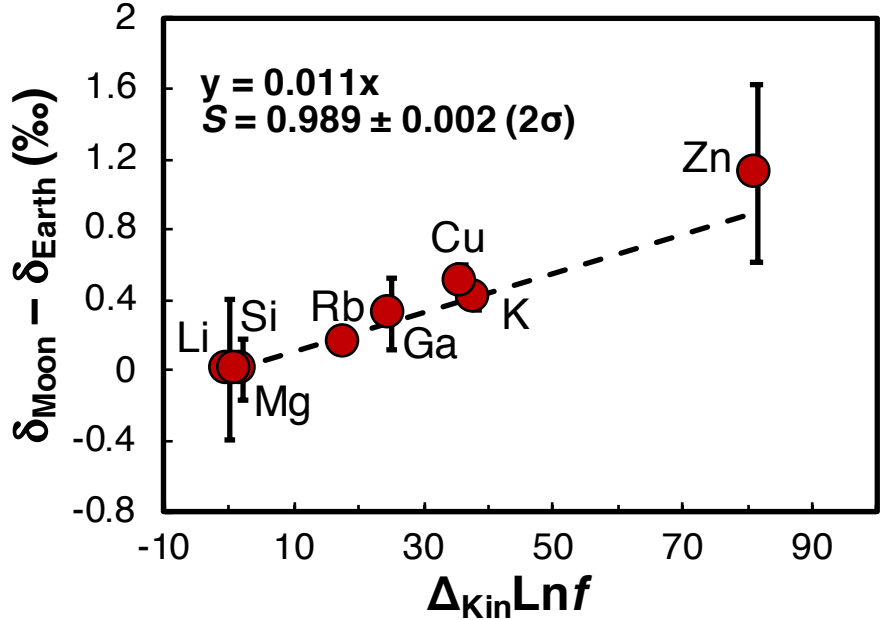

**Figure 2.** Assessment of the degree of saturation $S$ experienced by MVEs during their evaporation in the aftermath of the giant Moon-forming impact. Moon–Earth isotopic differences ($\delta_{Moon} - \delta_{Earth}$) are plotted against the product of kinetic isotope fractionation $\Delta_{kin}^{v-l} = [(m_j/m_i)^\beta - 1] \times 1000$ (with $m_j/m_i$ the ratio of the masses of the isotopes involved in defining $\delta^{i/j}$ and $\beta \simeq 0.5$ for most elements), and the natural logarithm of the fraction of the element remaining in the Moon ($\ln f$). According to Eq. 3, if the elements depleted in the Moon experienced evaporation under the same saturation, the data points should follow a straight line of slope $1 - S$. The elements indeed define a straight line, and a weighted linear regression gives a value for $S$ of $0.989 \pm 0.002$. The isotopic data plotted in this diagram are: Rb (87/85) and K (41/39) from Fig. 1; Li (7/6) from Magna et al. (2006); Si (30/28) from Armytage et al. (2012) and Fitoussi & Bourdon (2012); Mg (26/24) from Sedaghatpour et al. (2013); Ga (71/69) from Kato & Moynier (2017) and Wimpenny et al. (2019); Cu (65/63) from Herzog et al. (2009); Zn (66/64) from Kato et al. (2015). Note that highly refractory elements such as Ti (49/47) (Millet et al. 2016) and Ca (40/44) (Simon & DePaolo 2010) would plot at the origin (0, 0) and would follow the trend. Tin isotopes (124/116) seem to be lighter in lunar basalts compared to the bulk silicate Earth (Wang et al. 2019), but this element exists in three oxidation states in planets (0, 2+ and 4+), which complicates estimations of the Sn isotopic compositions of the bulk Earth and Moon. If confirmed, this light isotope enrichment of the Moon relative to Earth could reflect the presence of large equilibrium isotopic fractionation between vapor and liquid for Sn (Wang et al. 2019) and/or a higher evaporation coefficient for Sn compared to other MVEs (Sect. 4, Eq. 7). The concentration data used to calculate $f$ are from O'Neill (1991), Albarède et al. (2015) and Ringwood & Kesson (1977). The values of $\beta$ are 0.43 for K and Rb (Richter et al. 2011); 0.41 for Mg (Mendybaev et al. 2013), 0.3 for Si (Mendybaev et al. 2013), and 0.5 for all other elements.

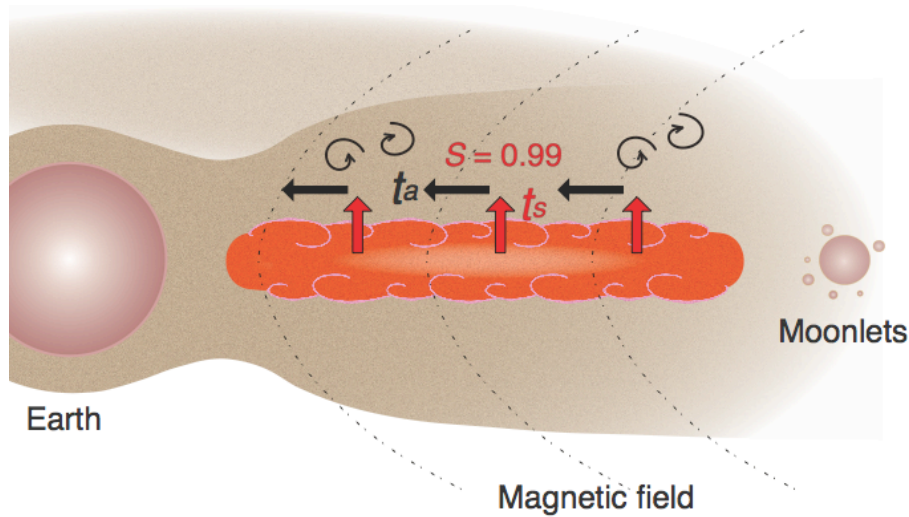

**Figure 3.** Possible setting for the depletion in moderately volatile elements of the Moon. The protolunar disk was composed of a liquid magma layer overlain by a vapor layer. The vapor was drained onto the Earth due to the turbulence and viscosity created by magnetorotational instability (MRI). The heavy isotopic compositions of the Moon compared to Earth for most MVEs (Figs. 1 and 2) can be explained by evaporation into a vapor medium that was slightly undersaturated ($S \simeq 0.99$). For this saturation to be achieved, the evaporation timescale ($t_s$) given by the mass of the vapor layer divided by the evaporation flux must have been balanced by the timescale for viscous accretion of the vapor layer onto Earth ($t_a$), and the viscosity of the vapor can thus be calculated. The vapor layer was convecting vigorously, and Kelvin–Helmoltz instabilities at the interface between the liquid and vapor layers prevented the development of a viscous layer.

**Table 1.** Rb Isotopic Compositions of Lunar and Terrestrial Rocks

|  | Rock Type | $\delta^{87}$Rb (‰) | 95% c.i. | La/U (ppm/ppm)[a] |
|---|---|---|---|---|
| **Lunar rocks** | | | | |
| 12002.613 | Olivine basalt | -0.034 | 0.036 | 29.6 |
| 12018.301 | Olivine basalt | 0.011 | 0.043 | 26.1 |
| 12052.353 | Pigeonite basalt | -0.003 | 0.036 | 18.5 |
| 10017.413 | Ilmenite basalt (high K) | 0.060 | 0.029 | 30.3 |
| 74275.361 | Ilmenite basalt | 0.089 | 0.058 | 45.3 |
| 77215.276 | Cataclastic norite (white-colored fragment) | -0.149 | 0.032 | 16.8 |
| **Terrestrial rocks** | | | | |
| BCR-2 | Basalt | -0.155 | 0.005 | 14.8 |
| BHVO-2 | Basalt | -0.114 | 0.006 | 35.7 |
| BE-N | Basalt | -0.103 | 0.018 | 34.2 |
| W-2 | Diabase | -0.147 | 0.014 | 18.9 |
| AGV-2 | Andesite | -0.146 | 0.012 | 20.2 |
| GSR-1 | Granite | -0.178 | 0.019 | 2.9 |
| GS-N | Granite | -0.177 | 0.014 | 10 |
| G-A | Granite | -0.234 | 0.020 | 8 |
| G-3 | Granite | -0.234 | 0.009 | 43 |

Notes.
[a]La/U ratios were calculated using concentration data from the Lunar Sample Compendium (https://curator.jsc.nasa.gov/lunar/lsc/), the USGS website (https://crustal.usgs.gov/geochemical_reference_standards/index.html), and Govindaraju (1994). For comparison, the La/U (ppm/ppm) ratio of CI chondrites is 27.6. c.i. is confidence interval.

**Acknowledgements.** Discussions with F.M. Richter, F. Ciesla, D.J. Stevenson, R.M. Canup, C.W. Visscher, S. Charnoz, S.J. Lock, J. Hu, H. Zeng, M. Meheut, M. Blanchard, J.Z. Zhang, S.M. Aarons, and A.W. Heard were greatly appreciated. We thank CAPTEM and R.A. Zeigler for providing the Apollo lunar samples analyzed in this study. This work was supported by a NASA NESSF fellowship (NNX15AQ97H) to NXN, NASA grants NNX17AE86G (LARS), NNX17AE87G (Emerging Worlds), and 80NSSC17K0744 (Habitable Worlds) to ND.

# Appendix
## A.1. Saturation Timescale

In this section, we calculate the timescale for replenishing a hydrostatic atmosphere above the disk midplane such that the surface pressure is at a set saturation $S$. The vertical forces acting on a parcel of gas are the pressure forces and the vertical component of the gravitational pull exerted by the Earth,

$$\frac{1}{\rho_g}\frac{\partial P}{\partial z} = -\frac{GM_\oplus z}{(R^2+z^2)^{3/2}}, \quad (8)$$

where $P$ is the gas pressure, $z$ is the vertical distance from the midplane, $\rho_g$ is the gas density, $G$ is the gravitational constant, $M_\oplus$ is the mass of the Earth, and $R$ is the distance from Earth's center. The gas density $\rho_g$ is related to the partial pressure $P$ through the ideal gas law and we have,

$$\frac{1}{P}\frac{\partial P}{\partial z} = -\frac{\bar{m}GM_\oplus z}{kT(R^2+z^2)^{3/2}}, \quad (9)$$

where $\bar{m}$ is the mean molecular mass of the vapor (~30 g mol⁻¹, equivalent to ~5× 10⁻²⁶ kg molecule⁻¹; Ward 2011), $k$ is Boltzmann's constant (1.38 × 10⁻²³ m² kg s⁻² K⁻¹), and $T$ is the temperature (in K). Integration of this differential equation with $P(z=0) = P_0$ yields,

$$P = P_0 e^{-\frac{\bar{m}GM_\oplus}{kTR}\left(1-\frac{1}{\sqrt{1+z^2/R^2}}\right)}. \quad (10)$$

With $z \ll R$ (the scale height is much smaller than the distance to the Earth), we have,

$$P = P_0 e^{-\frac{\bar{m}GM_\oplus}{2kTR^3}z^2}, \quad (11)$$

which can be rewritten as,

$$P = P_0 e^{-\frac{z^2}{2h^2}}, \quad (12)$$

with $h = \sqrt{\frac{kTR^3}{\bar{m}GM_\oplus}}$ the scale height of the vapor layer. Assuming that the vapor layer is well mixed, the scale height of all elements is the same and the partial pressure of an element $i$ as a function of altitude can be written as,

$$P_i = P_{i,0} e^{-\frac{z^2}{2h^2}}. \quad (13)$$

Introducing the sound speed $c = \sqrt{kT/\bar{m}}$ and the angular keplerian velocity $\Omega = \sqrt{GM_\oplus/R^3}$, the scale height can be rewritten as $h = c/\Omega$. We now evaluate what the timescale $t_{S,i}$ is for an element to be maintained at a given level of saturation $S_i$. The total surface density for an element (for the half-space; in kg m⁻²) is given by the integral,

$$\Sigma_i = \int_0^{+\infty} \frac{m_i P_i(z)}{kT} dz = \frac{m_i P_{i,0}}{kT} \int_0^{+\infty} e^{-\frac{z^2}{2h^2}} dz = \sqrt{\frac{\pi}{2}} \frac{h m_i P_{i,0}}{kT}, \quad (14)$$

where $m_i$ is the mass of the vapor species for element $i$ (in kg). The net mass flux of an element (in kg m⁻² s⁻¹) across the liquid/vapor interface is given by the Hertz-Knudsen equation (note that the more commonly used molar flux would have $m_i$ at the denominator),

$$\phi_i = \gamma_i \sqrt{\frac{m_i}{2\pi kT}} (P_{i,\text{eq}} - P_{i,0}), \quad (15)$$

where $\gamma_i$ is an evaporation coefficient. The saturation timescale $t_{S,i}$ is the total mass surface density of an element in the vapor layer at saturation $S_i$ (Eq. 14 with $P_{i,0} = S_i P_{i,\text{eq}}$) divided by the mass flux across the liquid/vapor layer at the same saturation given by the Hertz-Knudsen equation (Eq. 15, again with $P_{i,0} = S_i P_{i,\text{eq}}$),

$$t_{S,i} = \frac{\Sigma_i}{\phi_i} = \frac{\pi S_i}{\gamma_i(1-S_i)}\sqrt{\frac{m_i R^3}{\bar{m} G M_\oplus}}. \quad (16)$$

This saturation timescale is akin to a residence time.

*A.2. Chemical Purification of Rubidium*

A new Rb purification procedure was developed to separate Rb from rock matrices for isotopic analysis. The Rb blank of the procedure (digestion and column chemistry) is ~0.14 ng, which accounts for less than 0.5 % of total Rb from a typical sample (≥ 40 ng). The total Rb purification yields exceed 95 % for all the samples.

The separation of Rb from matrix elements is difficult because it is a trace element and it behaves very similarly to K, which is usually in much higher concentration (the K/Rb weight ratio of terrestrial rocks is ~325). Our procedure uses (i) AG50W-X8 cation exchange resin and $HNO_3$ to first separate Rb from matrix elements other than K, and (ii) Eichrom crown ether Sr resin (4,4′(5′)-di-t-butylcyclohexano 18-crown-6 in 1-octanol) and $HNO_3$ to then separate Rb from K. Previous Rb isotope studies (Pringle & Moynier 2017; Nebel et al. 2011) used mainly cation exchange resin and HCl to separate Rb from K and matrix elements. The second step in our procedure is similar to the chemistry used by Zhang et al. (2018), who used Sr resin but in a shorter column to separate Rb from K.

Samples of about 100 mg or less were digested in three steps using mixtures of concentrated $HF$-$HNO_3$-$HCl$-$HClO_4$ acids:

(i) 4 mL 28 M HF + 2 mL 15 M $HNO_3$ + 1 mL 10 M $HClO_4$ was added to each sample in a 30 mL fluoropolymer beaker that was closed and left on a hotplate at 130°C for 24 hours.

(ii) The solution was dried down at 130 °C on a hotplate, taken up in 4.5 mL 11 M HCl + 1.5 mL 15 M $HNO_3$, and left on a hotplate at 130 °C for 24 hours with the lid closed. This step was repeated twice.

(iii) After evaporation to dryness, the sample was re-dissolved in 4 mL 15 M $HNO_3$, and left on a hotplate for 24 hours with the lid closed. If a sample was not fully digested, the residue was transferred to a fluoropolymer beaker to be further digested in a Parr bomb at 175-180 °C for at least 3 days and it was then combined with the previously digested solution.

The combined solution was dried and re-dissolved in 1 M $HNO_3$ for column chemistry, which consists of five columns:

(i) The first column (Fig. 4(a)) uses 16 mL cation resin (AG50W-X8, 200-400 mesh resin in 20 mL Bio-Rad Econo-Pac columns of 15 mm diameter and 140 mm length). The samples were loaded in 4 mL 1 M $HNO_3$, and 160 mL of 1 M $HNO_3$ was then passed through the column to pre-purify Rb from most matrix elements (Strelow 1960; Strelow et al. 1965; Dybczyński 1972). This step removes elements with a higher distribution coefficient than Rb that are retained on the column (*e.g.*, Fe, Mg, Al, Ca, Cr, and Mn) while Rb is eluted. All the matrix elements with distribution coefficients lower than, or similar to Rb are collected together with Rb (1-160 mL; light-blue shaded area in Figure 4(a)). The elution scheme was designed to process relatively large sample masses (typically 100 mg). We collect all elution before the Rb peak to avoid Rb loss as excessive matrix element concentrations could potentially cause Rb to be eluted early by competing for ion exchange sites on the resin. This step removes a significant portion of matrix elements and all Rb is recovered.

(ii) The second column (Fig. 4(b)) uses the same type of resin (16 mL AG50W-X8 200-400 mesh in 20 mL Bio-Rad Econo-Pac columns of 15 mm diameter and 140 mm length) but a different $HNO_3$ molarity of 0.5 M. The reason for using this lower acid molarity is that the elements are better separated from each other (Strelow 1960; Strelow et al. 1965), the downside being a larger elution volume compared to the first column (the Rb peaks are at 280 *vs.* 120 mL respectively; Fig. 4). Samples were loaded in 4 mL of 0.5 M $HNO_3$, followed by 360 mL 0.5 M $HNO_3$ for elution. Rubidium is collected in 130-360 mL elution volume (light-blue shaded area in Fig. 4(b)). Elements that have a higher or lower partition coefficient than Rb are eliminated. Most Ti is removed in this step, while most K is kept in the Rb elution fraction.

(iii) After step (ii), a small amount of Ti may remain with Rb. The quantitative removal of remaining Ti was achieved by using a small anion column (1 mL resin of AG1-X8 200-400 mesh in a Bio-Rad Poly-Prep 10 mL column of 8 mm diameter and 90 mm length). The samples were loaded on the resin in 0.5 mL of 2 M HF and then eluted with 10 mL of the same acid. In 2 M HF, Ti sticks to the anion exchange resin while Rb is eluted (Faris 1960; Nelson et al. 1960).

(iv) This step separates Rb from K. We use a custom-made fluoropolymer column (40 cm in length and 0.4 cm in diameter) filled with Eichrom Sr resin (50-100 µm) to separate Rb from K. Element partitioning data for the Eichrom Sr resin can be found in Philip Horwitz et al. (1992). Samples were loaded and eluted in 3 M $HNO_3$. As shown in Fig. 4(c), complete separation of K and Rb was achieved. This step is similar to the one reported by Zhang et al. (2018) but we use a much longer column (40 cm *vs.* 13 cm) to better separate Rb from K.

(v) The solutions from the previous Sr resin step contained pure Rb but were relatively viscous, likely due to dissolution of small amounts of organics from the Sr resin. An additional small clean-up column (1 mL AG50W-X8 resin in a Bio-Rad Poly-Prep 10 mL column of 90 mm height and 8 mm diameter) was therefore used. Samples were loaded onto the column in 1 mL of 0.5 M $HNO_3$, followed by another 1 mL 0.5 M $HNO_3$, and Rb was then collected in 10 mL 6 M $HNO_3$. This step is not necessary for all Rb samples especially for samples with high Rb contents. Our tests showed that solutions before and after this column gave the same Rb isotopic compositions when measured by MC-ICPMS.

Pringle & Moynier (2017) used 0.5 M HCl and a 1 mL homemade column (0.6 cm inner diameter and 3.5 cm length) filled with AG50W-X8 resin to separate Rb from K. They claimed that this column could separate K from Rb but did not show any elution curve. We tested the method, and as shown in Fig. 5, the elution peaks of K and Rb largely overlap. In contrast, our method (Fig. 4(c)) allows a complete separation of Rb from K.

All purified Rb solutions were checked for the presence of potential matrix elements (the elements shown in Fig. 4) before measuring their Rb isotopic compositions. After the procedure, K was reduced to about the same concentration level as Rb or lower, and all other matrix elements had lower concentrations than Rb. The remaining K did not affect the Rb isotopic analyses. Our tests with K-doped Rb solutions showed that the measured isotopic composition of Rb was shifted only when K/Rb concentration ratios (ppm/ppm) were > 50 (Fig. 6).

Even small quantities of Sr in the Rb solution can produce a significant isobaric interference of $^{87}Sr^+$ on $^{87}Rb^+$. The contribution of $^{87}Sr$ was corrected for by monitoring $^{88}Sr$ and assuming a constant $^{87}Sr/^{88}Sr$ ratio of 0.085. Our tests with Sr-doped Rb solutions showed that solutions with $^{88}Sr/^{85}Rb$ intensity ratios (V/V) of ≤0.001 yielded accurate $\delta^{87}Rb$ isotopic compositions within ± 0.03 ‰ (Fig. 7). The $^{88}Sr/^{85}Rb$ intensity ratios (V/V) of our purified samples were all < 0.0005 (for most samples the ratio was much lower), which is ten times lower than the ratio after purification in Pringle & Moynier (2017) ($^{88}Sr/^{85}Rb$ < 0.005).

*A.3. Mass Spectrometry*

The Rb isotopic analyses were performed using the Thermo Scientific Neptune multi-collector inductively coupled plasma mass spectrometer (MC-ICPMS) at the University of Chicago. A standard-sample bracketing technique was used. Rubidium sample and standard solutions of ~ 15-25 ppb (~ 1-1.5 V for $^{85}Rb$ in low resolution) in 0.3 M $HNO_3$ medium were introduced into the MC-ICPMS at a flow rate of 100 µL/min via a dual cyclonic-Scott type quartz spray chamber. The samples and standards were matched for Rb concentrations within 1 % for MC-ICPMS analyses. Normal Ni sampler and skimmer cones were used, and all measurements were performed in low resolution mode. The MC-ICPMS at the University of Chicago is equipped with 9 Faraday collectors. Rubidium-85 and -87 were measured on L2 and axial (A) Faraday collectors respectively, and the isobaric interference of Sr was monitored at mass $^{88}Sr$ on H1. All three collectors were equipped with the $10^{11}$ Ω amplifiers. Isobaric interference $^{87}Sr$ was corrected for by assuming a constant $^{87}Sr/^{88}Sr$ ratio of 0.085, which is the terrestrial Sr ratio. For our purified samples, the correction of the $^{87}Sr$ interference on $^{87}Rb$ is small enough that uncertainties in the $^{87}Sr/^{88}Sr$ ratio are largely inconsequential (Fig. 7). For example, adopting two extreme $^{87}Sr/^{88}Sr$ values covering documented variations among chondrites, terrestrial, and lunar rocks (0.0835 and 0.0885, which correspond to $^{87}Sr/^{86}Sr$ ratios of 0.70 and 0.74, respectively) would shift $\delta^{87}Rb$ values of the most Sr-rich purified solutions ($^{88}Sr/^{85}Rb$ intensity ratio of ~ 0.0005) by 0.008 ‰ at most.

Data were collected as a single block of 25 cycles of 4.194 s integration time, with a take up time of 90 s and a washout time (0.45 M $HNO_3$ for wash) of 60 s. The clean acid solution (0.3 M $HNO_3$) that was used for diluting Rb for isotopic analysis was measured before and after each sample and standard under conditions identical to the sample or the standard (with the same integration, take up, and washout times). For each sample and standard, the average intensity of the two bracketing blank acids (0.3 M $HNO_3$) was subtracted. The Rb background was normally around 0.001-0.003 V for $^{85}Rb$ and increased by 0.001 V after a measurement session (typically ~15 hours or less).

The sample solutions were measured 5-12 times, depending on the total Rb amount, and the average $\delta^{87}Rb$ values were calculated. The uncertainty for a sample was calculated using the formula of $2 \times \sigma/\sqrt{n}$, with $n$ the number of replicates for the sample and σ the standard deviation of the $\delta^{87}Rb$ values of the standards calculated by treating the standards as if they were samples and calculating the $\delta^{87}Rb$ values by considering the 2 nearest standards bracketing each standard (Dauphas et al. 2009). All Rb isotopic compositions are expressed in $\delta^{87}Rb$ notation, which is the departure in ‰ of the $^{87}Rb/^{85}Rb$ ratio of a sample from that of the reference material NIST SRM984,

$$\delta^{87}\text{Rb (‰)} = [(^{87}\text{Rb}/^{85}\text{Rb})_{sample}/(^{87}\text{Rb}/^{85}\text{Rb})_{SRM984} -1] \times 1000.$$

*Test results*

The yield and accuracy of the procedure were checked with the reference standard NIST SRM984 treated as a sample, and with various synthetic and natural samples. Synthetic peridotite samples were made by mixing geostandard (peridotite) powders containing very little Rb with the Rb reference standard SRM984. The matrix elements in these synthetic samples are found in natural peridotites while their Rb isotopic compositions should be identical or very similar to SRM984 (*i.e.,* $\delta^{87}\text{Rb} = 0$). Two synthetic samples were made: DTS-2b (dunite) + SRM984 and PCC-1 (peridotite) + SRM984. as shown in Fig. 8, the two synthetic Rb-doped peridotite samples, and SRM984 treated as sample gave $\delta^{87}\text{Rb}$ values of zero within error.

We also measured several geostandards, including basalts and granites (BHVO-2, BCR-2, BE-N, W-2, AGV-2, GSR-1, GS-N, G-A and G-3). The majority of samples were digested and measured more than once and yielded reproducible results, which are in good agreement with previously published results (Pringle & Moynier 2017; Zhang et al. 2018) (Fig. 8). The Allende carbonaceous chondrite (powder from the Smithsonian Institution) was also measured to test the method on a low-Rb sample. Compared with the geostandards, Allende has much lower Rb concentration (~1 ppm *vs.* tens of ppm or more in the geostandards). Its Rb concentration is however similar to mare basalts (~1 ppm). It was digested and measured three times and all the three measurements yielded reproducible results (Fig. 8) Our measured Rb isotopic compositions of lunar samples in general agree with the data reported in Pringle and Moynier (2017) (Table 2).

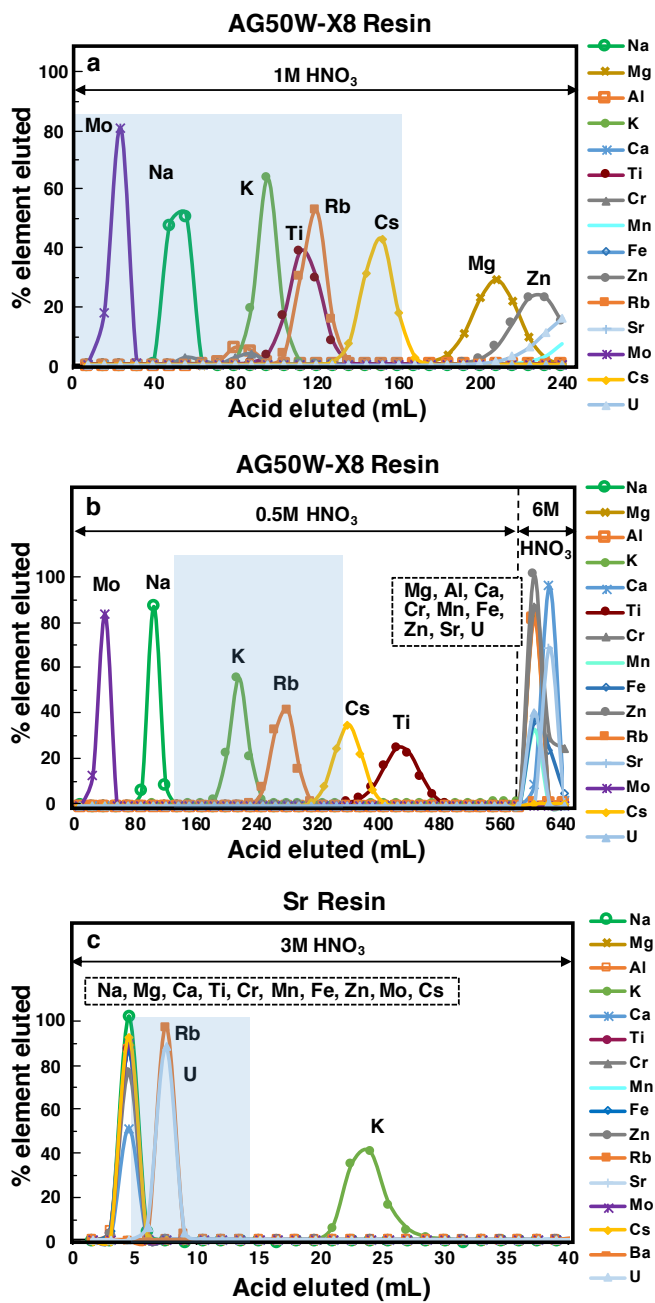

**Figure 4.** Rubidium separation scheme. The calibration curves were generated using synthetic solutions with equal concentrations of each element. In each panel, the x-axis is the cumulative volume of acids eluted in mL while the y-axis is the percentage of an element eluted. The light-blue shaded areas represent the elution cuts where Rb is collected. (a) Elution curves for various elements on 16 mL AG50W-X8 (200-400 mesh) cation resin in 1 M $HNO_3$. Elements not shown in the graph but present in the legend are retained on the column and most can be eluted with 6 M $HNO_3$. The first 160 mL of the elution is collected. (b) Elution using the same type of cation resin as in (a) but a lower acid molarity of 0.5 M. This step collects eluents around the Rb peak (130-360 mL) and removes almost all matrix elements except for K. (c) Separation of Rb from K using Eichrom Sr resin and 3 M $HNO_3$ (4-14 mL is collected). The resin separates Rb and K efficiently.

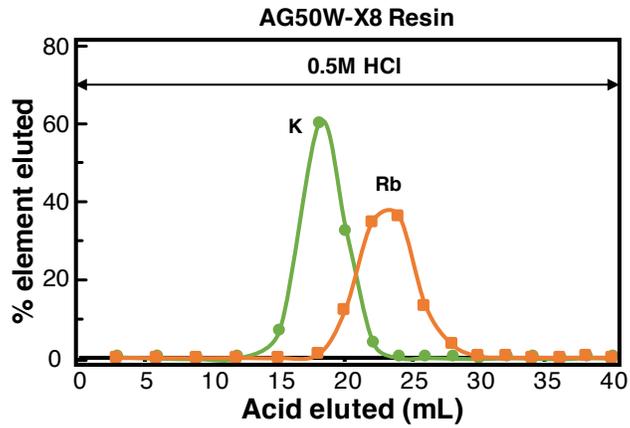

**Figure 5.** Elution curves of K and Rb using the separation method described in Pringle & Moynier (2017). In this method, a 1 mL column (0.6 cm inner diameter and 3.5 cm length) is filled with AG50W-X8 200-400 mesh resin and 0.5 M HCl is used to elute Rb and K. As shown, there is significant peak overlap between the Rb and K peaks and the two elements cannot be completely separated from each other (see panel c of Fig. 4 for a comparison with the procedure used in the present study).

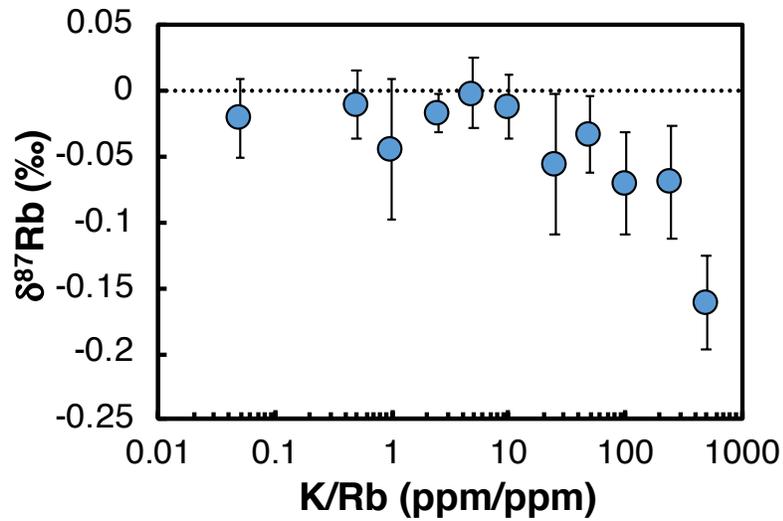

**Figure 6.** Influence of K on the isotopic composition of Rb. The measurements were performed using a spray chamber as the sample introduction system, in low resolution mode, using standard Ni cones, at a Rb concentration of ~20 ppb in 0.3 M $HNO_3$. The solutions were doped with various amounts of K. The isotopic composition of Rb is unaffected up to a K/Rb (ppm/ppm) ratio of 50. For reference, the K/Rb ratios (ppm/ppm) of terrestrial and lunar rocks are 325 and 450, respectively. The K/Rb ratio after chemical purification is always lower than 1.

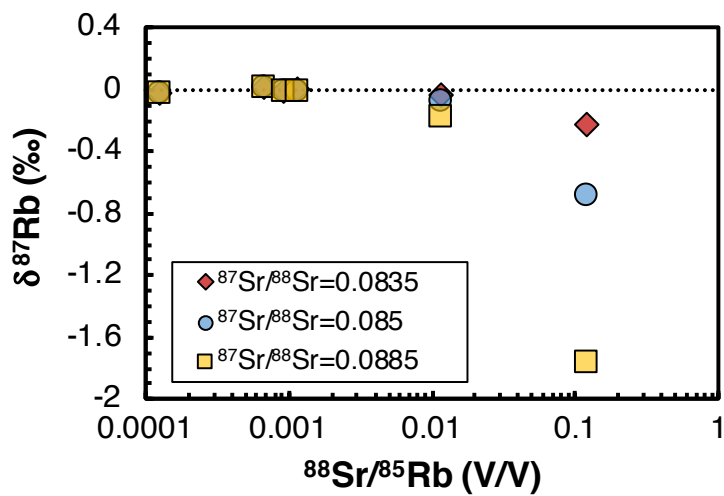

**Figure 7.** Influence of the Sr isobaric interference on the isotopic analysis of Rb. The measurements were performed using a spray chamber as the sample introduction system, in low resolution mode, using standard Ni cones, at a Rb concentration of ~20 ppb in 0.3 M $HNO_3$. The solutions were doped with various amounts of Sr. Strontium-88 was monitored for correcting the interference of $^{87}Sr$ on $^{87}Rb$ by assuming a constant $^{87}Sr/^{88}Sr$ ratio of 0.085 (blue circles). The correction is effective for $^{88}Sr/^{85}Rb$ intensity ratios (V/V) equal to or lower than 0.001. All the samples measured in this study have $^{88}Sr/^{85}Rb$ ratios (V/V) lower than 0.0005. Corrections of $^{87}Sr$ using different $^{87}Sr/^{88}Sr$ ratios of 0.0835 and 0.0885 (corresponding to $^{87}Sr/^{86}Sr$ ratios of 0.70 and 0.74) are also shown in red diamonds and yellow squares, respectively. For samples with $^{88}Sr/^{85}Rb$ ratios (V/V) below 0.001, the shifts in $\delta^{87}Rb$ values are smaller than 0.01 ‰.

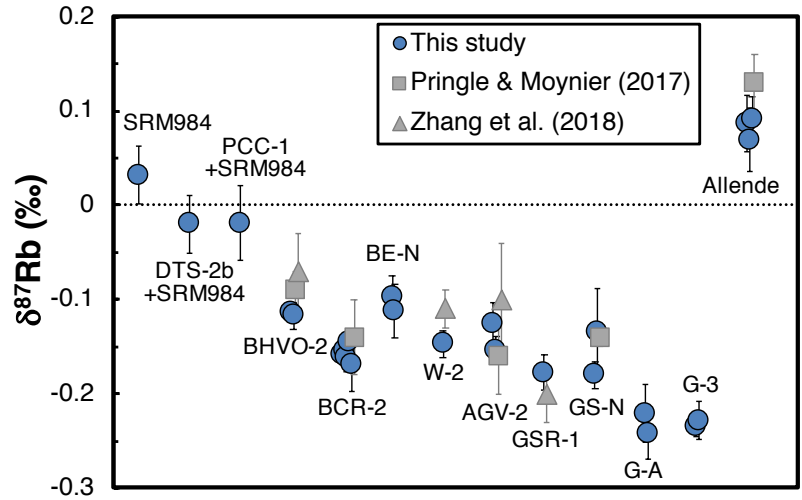

**Figure 8.** Test results for the reference standard SRM984 treated as a sample, synthetic samples (mixtures of DTS-2b+SRM984 and PCC-1+SRM984), geostandards (including basalts and granites) and carbonaceous chondrite Allende. SRM984 treated as a sample and the synthetic samples have Rb isotopic compositions identical to SRM984 within error. Geostandards and carbonaceous chondrite Allende yielded reproducible results (blue circles) that agree with literature data (grey squares and triangles).

**Table 2.** Compiled Literature Rb and K Isotopic Compositions and La/U Ratios of Lunar Samples

| Sample Identification | Rock Type | Isotopic Compositions | | La/U (ppm/ppm)[a] |
|---|---|---|---|---|
| **Rb isotopic compositions** | | $\delta^{87}$Rb (‰)[b] | 95% c.i.[b] | |
| 12012 | Olivine basalt | 0.08 | 0.01 | 26.7 |
| 15555 | Olivine-normative basalt | 0.07 | 0.03 | 20.7 |
| 12016 | Ilmenite basalt | 0.02 | 0.04 | 27.4 |
| 10003 | Ilmenite basalt (low K) | 0.14 | 0.05 | 54.3 |
| 10017 | Ilmenite basalt (high K) | -0.05 | 0.05 | 30.3 |
| 10057 | Ilmenite basalt (high K) | 0.06 | 0.06 | 33.7 |
| 77215 | Cataclastic norite | 0.02 | 0.04 | 16.8 |
| **K isotopic compositions** | | $\delta^{41}$K (‰)[c] | 95% c.i.[c] | |
| 10017.282 | Ilmenite basalt (high K) | 0.37 | 0.06 | 30.3 |
| 10071.126 | Ilmenite basalt (high K) | 0.46 | 0.04 | 39 |
| 14301.290 | Regolith breccia | 0.35 | 0.04 | 20.9 |
| 14305.330 | Crystalline matrix breccia | 0.42 | 0.04 | 22.2 |
| 60315.191 | Poikilitic impact melt | 0.42 | 0.05 | 23.3 |
| 12013.170 | Breccia with granite | 0.47 | 0.03 | 6.7 |
| 12013.171 | Breccia with granite | 0.60 | 0.05 | 6.7 |

Notes.
[a]La/U ratios were calculated using concentration data from the Lunar Sample Compendium (https://curator.jsc.nasa.gov/lunar/lsc/). For comparison, the La/U (ppm/ppm) ratio of CI chondrites is 27.6.
[b]Data from Pringle & Moynier (2017).
[c]Data from Wang & Jacobsen (2016).